\documentstyle[aps]{revtex}\tighten
\input epsf
\begin{document}

\title{Gamma-ray bursts as the birth-cries of black holes\thanks{This
essay received an ``honorable mention" in the Annual Essay
Competition of the Gravity Research Foundation for the year
2000.}}

\author{Pankaj S. Joshi$^1$\thanks{psj@tifr.res.in},
Naresh K. Dadhich$^2$\thanks{nkd@iucaa.ernet.in} and Roy
Maartens$^3$\thanks{roy.maartens@port.ac.uk}}

\address{~}

\address{$^1$Tata Institute of Fundamental Research, Mumbai~400~005, India}

\address{$^2$Inter-University Centre for Astronomy and Astrophysics,
Ganeshkind, Pune~411~007, India}

\address{$^3$Relativity and Cosmology Group, School of Computer Science
and Mathematics, Portsmouth University, Portsmouth~PO1~2EG, Britain}

\maketitle

\begin{abstract}

The origin of cosmic gamma-ray bursts remains one of the most
intriguing puzzles in astronomy. We suggest that purely general
relativistic effects in the collapse of massive stars could
account for these bursts. The late formation of closed trapped
surfaces can occur naturally, allowing the escape of huge energy
from curvature-generated fireballs, before these are hidden within
a black hole.

\end{abstract}

\[ \]

Gamma-ray bursts (GRBs) are the fireworks display of the cosmos.
They originate from super-high energy events ($10^{51}-10^{54}$
ergs), probably in star-forming regions, and appear to be
isotropically distributed and of cosmological origin \cite{exp}.
According to the fireball model \cite{fb}, GRBs are produced in
shocks (probably internal) in a relativistically expanding
fireball. Causality restricts the progenitor size to $\lesssim
3000$ km (up to $10^{12}$ cm in the fireball frame).

Speculations on the origin of GRBs include the capture of neutron
stars by black holes, the merger of binary neutron stars \cite{bh}
and the hypernova model \cite{hyp}. In both the latter models, an
accretion torus or disc forms around a rotating black hole, with
energy coming via a $10^{15}$ G magnetic field, or the
gravitational binding energy of the torus. Challenges for these
models include explaining the required stability of the torus, and
the required low baryon loading.

Considering the extreme nature of the energetics and durations,
and severe problems faced by almost all current models, we should
be open-minded about the possible nature of the progenitor. We
point out that gravitational collapse of a massive star, which has
exhausted its nuclear fuel, offers a rather natural GRB mechanism.
Our model is very much within the standard framework of general
relativity, based on certain very fundamental aspects, and less
speculative than some other proposals.

As stellar collapse progresses, gravity becomes so strong that
closed trapped surfaces start developing within the
interior of the star. These are surfaces from which both outgoing
and ingoing wave fronts converge to the centre of the cloud, and
no light escapes. The well-known singularity theorems then predict that
a spacetime singularity must develop. These theorems, however, do
{\it not} provide any information on the nature of such a
singularity, its physical properties and its possible interaction
with outside observers.

To address these questions, one needs to study the dynamics of
collapse of matter clouds as governed by the Einstein equations.
This has been done in considerable detail in the past decade or so
\cite{coll}, for type~I matter fields, which include all
physically reasonable observed matter. The generic conclusions
emerging from these studies are most striking: {\em while the
collapse always produces diverging curvature and density, trapped
surfaces may not develop early enough to shield the whole process
from outside.}

To be specific, consider a spherically symmetric matter cloud. The
interior is governed by Einstein's equations, and the exterior is
matched to Schwarzschild or Vaidya spacetime. The interior metric
in comoving coordinates is
\[
ds^2 = - e^{2\nu(t,r)} dt^2 + e^{2\psi(t,r)} dr^2 + R^2(t,r)
d\Omega^2\,,
\]
and the type~I stress-energy tensor is
\[
T^t_t=-\rho,\; T^r_r= p_r,\; T^\theta_\theta=T^\phi_\phi=
p_\theta,\; T^t_r=T^r_t=0\,,
\]
where $\rho, p_r$, and $p_\theta$ are the density, radial and
tangential pressures.

Initial data are the $t=0$ values of the three metric functions,
the density and the pressures (see \cite{ivp}). Integrating
Einstein's equations produces a free function $F(t,r)$, the total
mass within the shell of comoving radius $r$ at epoch $t$. Most
studies so far on gravitational collapse, involving dust, perfect
fluids, self-similar collapse, and so on, are part of this general
scenario, and form special subcases. As collapse proceeds, the
mass-energy density grows without bound. The development of
trapped surfaces can be traced explicitly via outgoing null
geodesics, and the equation for the trapped surface or {\it
apparent horizon} is given by
\begin{equation}\label{1}
F = R(t,r)\,.
\end{equation}
Within a finite time $t=t_s$, general relativity predicts that the
density and curvature scalars such as $R^{abcd}R_{abcd}$ diverge
to infinity. However, at the Planck scale $10^{-33}$ cm, quantum
gravity effects should take over, smearing out the classical
singularity. Hence, based on our current theoretical understanding
of gravity, we conclude that the end-product of collapse of a
massive star will be a super-dense region with extremely high
spacetime curvature and density---a curvature-generated fireball.
The key question is: what is the possible interaction of this
object with outside observers?

If the apparent horizon as governed by Eq. (\ref{1}) starts
developing earlier than the epoch of formation of the fireball,
then the event horizon will fully cover the fireball, which will
thus be hidden within a black hole. On the other hand, if trapped
surfaces form {\em later}, then it is possible for the fireball to
communicate with far-away observers (see Fig. 1). This means that
the physical effects caused by the fireball, and those created
within the immediately surrounding medium, will be able to
propagate into the universe. While a fireball always develops in
gravitational collapse, its communication properties depend on the
nature of the (regular) initial data. The properties of the
initial data determine whether the apparent horizon starts
developing early enough to shield the fireball, or develops at or
later than the fireball epoch.

\begin{figure}[h]
\parbox[b]{15.88cm}
{ \epsfxsize=15.80cm \epsfbox{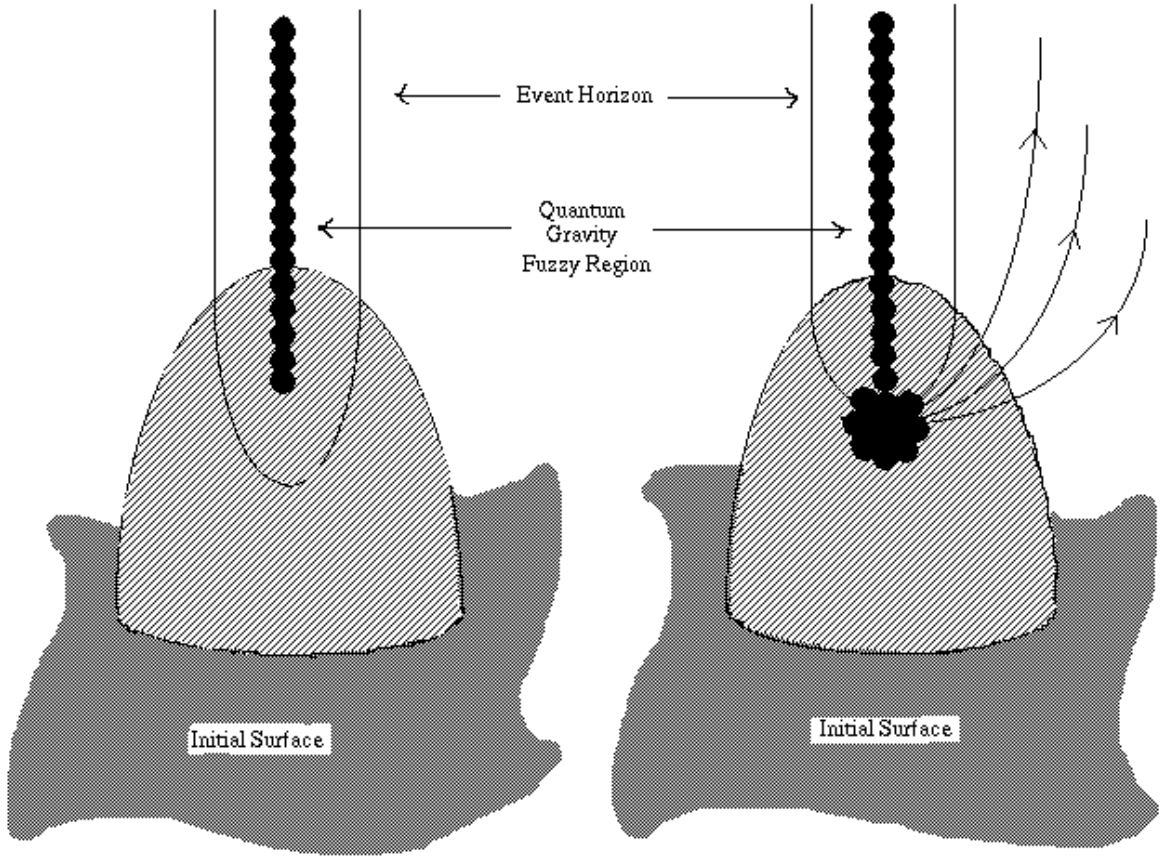} } \caption{ Formation
of black hole (left) versus a visible fireball (right) in
gravitational collapse.}
\end{figure}

To take a specific example, consider the collapse of a dust ball
\cite{dust}. When the dust is homogeneous, the apparent horizon
starts developing early enough, and the fireball is fully hidden
within a black hole. Now consider inhomogeneous dust with a
generic mass profile
\[
F(r)= F_0r^3 + F_1 r^5 + F_2 r^6 +\cdots\,,
\]
and with a peak at the centre, where the density gradient is
$\rho_n=(n+3)n!F_n$, decreasing outwards (which would be a
physically more realistic situation for any star). Then trapped
surface development is delayed, and given by
\[
t_{ah}(r) = {2\over3} F_0^{-1/2} - {F_nr^n\over 3F_0^{3/2}} -
{2\over3} F_0r^3\,.
\]
Thus the fireball (where $\rho\to\infty$,
$R^{abcd}R_{abcd}\to\infty$), together with its surrounding
medium, is able to communicate with far-away observers.

This meets the necessary basic criteria of the fireball proposal
\cite{fb}. GRBs are created by the shocks formed in the
surrounding medium by the fireball. The important point is that we
have a natural and robust mechanism for the creation of the
fireball, within the {\em standard} and familiar framework of
general relativity, without invoking additional or exotic
mechanisms.

Our mechanism is dominated by classical astrophysical effects and
does not rely on quantum particle production (see \cite{cj}),
although this will enhance the process. Also, our conclusions do
not depend on any specific equation of state. This allows for the
surrounding medium to become radiation-dominated, which is
expected, so that it expands relativistically (there are no static
solutions available to confine such a fluid). This will again
create relativistic expansion and shocks, which will generate the
GRBs as in the currently available mechanisms.

The really crucial step is the creation of a relativistic
fireball, and its communicability with far-away observers, which
depends on the initial data parameters from which the collapse
evolves. The advantage of such a model is that we do not really
need to see the rays coming {\it from} the fireball itself, which
need not be visible. The actual role of the fireball, which is an
envelope with a Planck-size super-strong curvature core and
surrounding medium, is to create shocks in the matter outside via
its relativistic expansion, which in turn gives rise to GRBs,
before the fireball disappears within the black hole forever. For
the collapse scenario with general matter fields as considered
above, it can be established rigorously that the fireball will
disappear within a black hole eventually. This feature naturally
accounts for finite-duration bursts. In this sense, we may
describe GRBs as birth cries of black holes.

From a physical viewpoint, the essential difference from a black
hole model for GRBs (which has many difficulties), is that in our
case, regions of arbitrarily high curvature are visible to a
distant observer, in principle. Physical processes, including
quantum particle creation, in these high-curvature regions then
cause the star to appear very different from a black hole, and
generate GRBs. In the black hole model, the inner part of the star
must necessarily form a black hole first, followed at later stages
by the infall of outer regions into the black hole. The required
energy for the burst is extracted during this infall, by a
suitable mechanism. In our model, the collapsing star develops
regions of very high curvature which are not hidden behind a
horizon. The key point is that a `window' becomes available for
escape of matter and energy from the extremely strong gravity
regions to far-away regions.

We have already emphasized that the conclusions on the development
of the fireball remain invariant, independent of the form of the
matter and equation of state used in collapse. One could also
inquire about stability and non-spherical perturbations. There
have been some studies on this \cite{rot}, and the indications are
that departures from spherical symmetry will not necessarily
change these conclusions, and may in fact preserve them.

The duration of GRBs can also be explained naturally within our
framework. The development of the fireball initiates just before
$t=t_s$. For a collapsing star of a few solar masses, the collapse
time-scale will be a few milliseconds in the comoving frame. As
soon as the fireball is created, it generates shocks in the
surrounding medium, and because of the causal connection
available, the rays proceed outwards, coming out at $r = 2m +
\delta$ at the boundary of the star, where $\delta$ is a small
positive quantity (assuming a Schwarzschild exterior). This is the
beginning of the GRB for an outside observer. Very soon the
fireball disappears within the black hole, and the later part of
radiation starts coming from closer to the event horizon,
eventually being infinitely redshifted. Thus the outside observer
sees a GRB coming out, with a peak, and then a dying intensity.

\end{document}